\documentclass [12pt,a4paper]{article}

\title{Thermodynamics of Expansive Nondecelerative Universe}
\author{Miroslav S\'{u}ken\'{\i}k and Jozef \v{S}ima \\[1ex]
Slovak Technical University, Faculty of Chemical Technology\\
Radlinsk\'{e}ho 9, SK-812 37 Bratislava, Slovakia\\
e-mail: sukenik@minv.sk, sima@chtf.stuba.sk}
\date{}
\begin{document}
\maketitle
\begin{abstract}
  The present contribution deals with thermodynamic aspects of the
  model of Expansive Nondecelerative Universe. In this model, in the
  matter era a dependence $T_{CBR} \approx E_{CBR} \approx a^{ - 3 /
    4}$ holds for the energy of cosmic background radiation, $E_{CBR}
  $ and its temperature, $T_{CBR} $, while the proportionality of the
  energy density $\varepsilon _{CBR} $ to the gauge factor $a$ can be
  expressed as $\varepsilon _{CBR} \approx a^{-3}$. The given
  relationships comply with experimental observations of the cosmic
  background radiation as well as with a surprising finding that the
  Universe expansion is not decelerated by gravitational forces. It is
  rationalized that the specific entropy is proportional to
  $a^{-1/4}$, i.e.\ it is gradually decreasing in time.
\end{abstract}

\section{Introduction}

In the model of Expansive Nondecelerative Universe (hereinafter ENU) it 
holds \cite{ska91,ska93}
\begin{equation}
\label{eq1}
a = c.t_{c} = {\frac{{2G.m_{U}} }{{c^2}}}
\end{equation}
where $a$ is the gauge factor, $t_{c} $ is the cosmological time, and
$m_{U} $ is the mass of the Universe. The present-time values provided
by calculations based on ENU do not differ substantially from the
generally accepted values and are as follows: $a_{pt} \cong 1.3\times
10^{26}$ m, $t_{c(pt)} \cong 1.4\times 10^{10}$yr, $m_{U(pt)} \cong
8.6\times 10^{52}$ kg. It stems from~(\ref{eq1}) that the Universe
mass is time-increasing.  Since the mass-energy of the Universe must
be time-independent (and equal to zero), simultaneously with the
matter creation also an equivalent amount of gravitational field
energy is formed which is, however, negative. This is why the Universe
can expand with a constant velocity equal to the velocity of light $c$
in ENU. In ENU, Schwarzschild metric is replaced by Vaidya metric
\cite{vai51} originally elaborated to solve the problems of radiating
stars, latter shown by Virbhadra \cite{vir99,vir92} and by us
\cite{suk0101026} that its application is more general.

It is postulated in ENU that the energy density of the Universe is just 
critical and it is expressed as
\begin{equation}
\label{eq2}
\varepsilon _{crit} = {\frac{{3  c^{4}}}{{8\pi .G.a^{2}}}}
\end{equation}

Till the end of the radiation era, there was a thermodynamic equilibrium of 
matter and radiation, and energy, temperature and gauge factor were related 
as follows
\begin{equation}
\label{eq3}
E_{CBR} \approx T_{CBR} \approx a^{ - 1 / 2}
\end{equation}

The fact that the energy density in~(\ref{eq2}) is proportional to $a^{ - 2}$ and 
not to $a^{ - 3}$ can be rationalized by matter creation. Thus, ENU 
describes the Universe in which eternal inflation occurs. In classical 
inflationary models of universe, after completing its inflation stage the 
Universe should decelerate due to effects of gravitational forces. As a 
consequence, in the models of inflationary universe a new matter incessantly 
emerges from behind the event horizon and in this way the proportionality 
$\varepsilon \approx a^{ - 2}$ stated by~(\ref{eq2}) is explained. 

In the ENU model relation~(\ref{eq2}) holds permanently, i.e.\ also in
the matter era and the energy density matches well with the accepted
gauge factor value.  Detailed and precise observations performed in
the last few years have led to a conclusion that the predicted
decrease in the Universe expansion rate has not occurred. Contrary, a
nonzero value of the cosmological constant $\Lambda $ or a newly
elaborated quintessential model \cite{hue0104006} gave rise to a
presumption stated that the Universe expansion accelerates and that
such an acceleration started at the beginning of the matter era. The
postulate on the Universe acceleration leads, however, directly to a
very important conclusion concerning impossibility of
relation~(\ref{eq2}) validity in the matter era and also to a
conclusion on impossibility of critical energy density preserving.

\section{Thermodynamics of ENU}

It is generally accepted that the radiation era ended approximately in the 
time 

\begin{equation}
  \label{eq:4a}
t_{r} \cong 7\times 10^{5} \mbox{ yr}
\end{equation}
when the temperature of radiation approached to
\begin{equation}
  \label{eq:5a}
  T_{r} \cong 5\times 10^{3} \mbox{ K}  
\end{equation}
(In the contribution, the subscripts $pt$, $r$ and $m$ refer to the 
present-time, the end of radiation era, and matter era, respectively). The 
present-time temperature is
\begin{equation}
  \label{eq:6a}
T_{pt} \cong 2.735 \mbox{ K}   
\end{equation}
Taking into account that the Universe expansion did not decelerate in the 
matter era, a presumption emerges stating that not only event horizon but 
also original part of the Universe extended in about four orders
\begin{equation}
\label{eq4}
{\frac{{t_{pt}} }{{t_{r}} }} = {\frac{{a_{pt}} }{{a_{r}} }} \cong 10^{4}
\end{equation}

Based on the fact that from the end of the radiation era the of cosmic 
background radiation temperature has decreased by three orders but the gauge 
factor increased by four orders, a relation between the energy of cosmic 
background radiation $E_{CBR} $ and its temperature $T_{CBR} $ follows
\begin{equation}
\label{eq5}
T_{CBR} \approx E_{CBR} \approx a^{ - 3 / 4}
\end{equation}

Introducing the value of $a_{pt} $ into calculation of the present-time 
critical energy density of the Universe it follows that 
\begin{equation}
  \label{eq:9a}
  \varepsilon _{crit(pt)} \cong 8.577\times 10^{ - 10} \mbox{J/m$^{\rm 3}$ }
\end{equation}

The energy density of radiation can be extracted from Stefan-Boltzmann law 
and for cosmic background radiation is generally given as
\begin{equation}
\label{eq6}
\varepsilon _{CBR} = {\frac{{4\sigma .T^{4}}}{{c}}}
\end{equation}
Based on~(\ref{eq:6a}) and~(\ref{eq6}) the present-time energy density value of cosmic 
background radiation reaches
\begin{equation}
  \label{eq:11a}
  \varepsilon _{CBR(pt)} \cong 4.229\times 10^{ - 14} \mbox{ J/m$^{{\rm 3}}$} 
\end{equation}
It follows from~(\ref{eq5}) and~(\ref{eq6}) that during the matter era
\begin{equation}
\label{eq7}
\varepsilon _{CBR(m)} \approx a^{ - 3}
\end{equation}
and at the same time, stemming from~(\ref{eq2}),~(\ref{eq:9a}),
(\ref{eq6}),~(\ref{eq:11a}) and~(\ref{eq7}) it follows
\begin{equation}
\label{eq8}
{\frac{{a_{pt}} }{{a_{r}} }} = {\frac{{\varepsilon _{crit(pt)} 
}}{{\varepsilon _{CBR(pt)}} }}
\end{equation}
Based on~(\ref{eq8}) the gauge factor value at the end of the radiation era can be 
calculated 
\begin{equation}
  \label{eq:14a}
  a_{r} \cong 6.41\times 10^{21} \mbox{ m}
\end{equation}
together with the cosmological time value at the same time
\begin{equation}
  \label{eq:15a}
  t_{r} \cong 6.6\times 10^{5} \mbox{ yr}
\end{equation}
Treatment of relations~(\ref{eq:6a}),~(\ref{eq5}) and~(\ref{eq:14a}) leads to 
\begin{equation}
\label{eq9}
{\frac{{T_{pt}} }{{T_{r}} }} = \left( {{\frac{{a_{r}} }{{a_{pt}} }}} 
\right)^{3 / 4}
\end{equation}
Stemming from~(\ref{eq9}) the temperature at the end of the radiation era is 
directly calculable and it reaches the value of
\begin{equation}
  \label{eq:17a}
  T_{e} \cong 4650 \mbox{ K}
\end{equation}
being in excellent agreement with the generally accepted value obtained 
using other independent modes of calculations.

\section{Specific Entropy}

Total average number of relict photons $n(h\nu )_{m} $ in a cubic meter 
during the matter era relates to the gauge factor according to~(\ref{eq5}) and~(\ref{eq7}) 
as follows 
\begin{equation}
\label{eq10}
n(h\nu )_{m} \approx a^{ - 9 / 4}
\end{equation}
and that of protons $n(p)_{m} $ (representing the matter particles) based on 
(\ref{eq2}) as
\begin{equation}
\label{eq11}
n(p)_{m} \approx a^{ - 2}
\end{equation}

Dependence of the specific entropy $S$, defined as a number of relict 
photons per one proton, on the gauge factor is in the matter era expressed 
as 
\begin{equation}
\label{eq12}
S \approx a^{ - 1 / 4}
\end{equation}

At the time being, the temperature of cosmic background radiation (2.735 K) 
leads to the following number of relict photons in a volume unit
\begin{equation}
\label{eq13}
n(h\nu )_{pt} = {\frac{{\varepsilon _{CBR(pt)}} }{{E_{CBR(pt)}} }} \cong 
4\times 10^{8}
\end{equation}
where $E_{CBR(pt)} $ is the mean energy of actual relict photons.

Given the present time energy density~(\ref{eq2}) and gauge factor values, a number 
of protons in a volume unit reaches
\begin{equation}
\label{eq14}
n(p)_{pt} \cong 5
\end{equation}
The present-time specific entropy calculated as a ratio of values provided 
in~(\ref{eq13}) and~(\ref{eq14}) is of the order
\begin{equation}
\label{eq15}
S_{pt} \cong 10^{8}
\end{equation}
At the end of the radiation era, the specific entropy value approached
\begin{equation}
\label{eq16}
S_{r} \cong 10^{9}
\end{equation}
Comparison of~(\ref{eq15}) and~(\ref{eq16}) verifies the correctness of~(\ref{eq12}), i.e.\ a slow 
decrease in specific entropy with time.

Within a discussion on a time-dependence of specific entropy some 
contradictions emerge. If the specific entropy is constant, i.e.\ if relation 
(\ref{eq16}) is valid at the present-time number of relict photons and gauge factor, 
the Universe density would have to have subcritical value. The assumption of 
permanent critical (nearly critical) density, however, excludes a constant 
value of specific entropy. The majority of current cosmological models take, 
however, critical mass-energy density a priori into account and tries to 
solve this discrepancy introducing some ``exotic'' nonbaryonic forms of 
matter.

\section{Thermodynamics and Gravitation}

In the period of time starting with the Universe expansion and finishing 
with the end of the radiation era, energy densities given by~(\ref{eq2}) and~(\ref{eq6}) 
are identical which corresponds to~(\ref{eq3}). It must therefore hold for the mean 
energy value of the photons of cosmic background radiation 
\begin{equation}
\label{eq17}
E_{CBR} = E_{Pc} \left( {{\frac{{l_{Pc}} }{{a}}}} \right)^{1 / 2} = \left( 
{{\frac{{\hbar ^{3}.c^{7}}}{{G.a^{2}}}}} \right)^{1 / 4}
\end{equation}
where $l_{Pc} $ and $E_{Pc} $ are the Planck length and Planck energy \cite{gro00}, 
respectively

\begin{eqnarray}
  l_{Pc} &=& \left( {{\frac{{G.\hbar} }{{c^{3}}}}} \right)^{1 / 2} = 
1.606151\times 10^{ - 35} \mbox{ m} \label{eq:26a} \\
  E_{Pc} &=& \left( {{\frac{{\hbar .c^{5}}}{{G}}}} \right)^{1 / 2} = 
1.2211\times 10^{19} \label{eq:27a}
\end{eqnarray}
Since proportionality~(\ref{eq5}) holds in the matter era, the mean energy value of 
the photons of relict radiation is expressed as
\begin{equation}
\label{eq18}
E_{CBR} = \left( {{\frac{{\hbar ^{3}.c^{7}.a_{r}} }{{G.a^{3}}}}} \right)^{1 
/ 4}
\end{equation}

The ENU model allows to localize gravitational field energy. The 
wave function of gravitational field is described \cite{suk0010061} as
\begin{equation}
\label{eq19}
\Psi _{g} = \exp {\left[ {i.t\left( {{\frac{{m.c^{5}}}{{\hbar .a.r^{2}}}}} 
\right)^{1 / 4}} \right]}
\end{equation}
where $\Psi _{g} $ is the wave function of gravitational field quanta created 
by a body with the mass $m$ at the distance $r$. The mentioned thermodynamic 
equilibrium in the radiation era means that the total mass of the relict 
radiation is equal to the total mass of matter particles. This is why this 
mass can be expressed as 
\begin{equation}
\label{eq20}
m \cong {\frac{{a.c^{2}}}{{G}}}
\end{equation}
When taking the Universe as one system into account, the following general 
equation must always hold
\begin{equation}
\label{eq21}
r = a
\end{equation}
Introducing~(\ref{eq20}) and~(\ref{eq21}) into~(\ref{eq19}), we obtain the Universe wave function in 
the form \cite{suk0010061}
\begin{equation}
\label{eq22}
\Psi _{g} = \exp {\left[ {i.t\left( {t_{Pc} .t_{c}}  \right)^{ - 1 / 2}} 
\right]}
\end{equation}

Here we can formulate a hypothesis stating that the mean energy of a photon 
of relict radiation is modulated by the energy of gravitational quanta 
$E_{g} $and we will intent to justify this hypothesis and prove its 
correctness. In such a case the absolute values of the corresponding mean 
energies must be identical, i.e.
\begin{equation}
\label{eq23}
{\left| {E_{g}}  \right|} = E_{CBR} 
\end{equation}
Writing a Schr\"{o}dinger-like equation for the energy of gravitational 
quanta $E_{g} $
\begin{equation}
\label{eq24}
E_{g} .\Psi _{g} = i.\hbar {\frac{{d\Psi _{g}} }{{dt}}}
\end{equation}
it comes from~(\ref{eq22}) and~(\ref{eq24}) that
\begin{equation}
\label{eq25}
{\left| {E_{g}}  \right|} = {\frac{{\hbar} }{{\left( {t_{Pc} .t_{c}}  
\right)^{1 / 2}}}} = \left( {{\frac{{\hbar ^{3}.c^{7}}}{{G.a^{2}}}}} 
\right)^{1 / 4}
\end{equation}
which is a relation identical to that of~(\ref{eq17}).

Situation is quite different in the matter era. Stemming from the validity 
of~(\ref{eq7}) for the total mass of the photons of relict radiation, 
$m_{CBR(total)} $ it must hold
\begin{equation}
\label{eq26}
m_{CBR(total)} \approx V.a^{ - 3} = \mbox{const}.
\end{equation}
At the same time
\begin{equation}
  \label{eq:37a}
  m_{CBR(total)} \cong {\frac{{a_{r} .c^{2}}}{{G}}} \cong 10^{49}
  \mbox{ kg}
\end{equation}
This value is really constant. At the present-time volume of the Universe 
being
\begin{equation}
  \label{eq:38a}
  V_{U} \cong 10^{79} \mbox{ m$^{{\rm 3}} $}
\end{equation}
and the energy density of relict radiation~(\ref{eq:11a}) one can easily come, in an 
independent way, to the mass given by~(\ref{eq:37a}). This mass generates the 
gravitational field that modulates the mean value of the energy of relict 
photons in the matter era. This allows to introduce~(\ref{eq21}) and (\ref{eq:37a}) into~(\ref{eq19}) 
and obtain
\begin{equation}
\label{eq27}
\Psi _{g} = \exp {\left[ {i.t\left( {{\frac{{a_{r} .c^{7}}}{{\hbar 
.G.a^{3}}}}} \right)^{1 / 4}} \right]}
\end{equation}
The mean energy of photons in the matter era, stemming from~(\ref{eq24}) and~(\ref{eq27}) is
\begin{equation}
\label{eq28}
{\left| {E_{g}}  \right|} = E_{CBR} = \left( {{\frac{{\hbar ^{3}.c^{7}.a_{r} 
}}{{G.a^{3}}}}} \right)^{1 / 4}
\end{equation}

Relation~(\ref{eq28}) is identical to postulated relation~(\ref{eq18}). In this way, using 
the ENU model we are able to harmonize theory and observation. It holds that 
the mean energy of the photons of relict radiation is equal to the absolute 
value of the energy of gravitational quanta that are generated by the total 
mass of relict radiation $m_{CBR(total)} $.

\section{Conclusions}

All values presented in this contribution comply with experimentally 
observed or generally accepted values.

Stemming from a supposed acceleration of the Universe expansion hypothesed 
in the models including nonzero value of cosmological constant $\Lambda $ in 
those introducing quintessence (or other ``exotic'' forms of matter and 
energy), values of the Universe-related quantities might differ in a 
substantial extent in the future. Some of the open questions can be answered 
by exact measurements of the parameter $\omega $ (defined as the 
pressure-to-energy density ratio) 
\begin{equation}
\label{eq29}
\omega = {\frac{{p}}{{\varepsilon} }}
\end{equation}
that in the models accepting the nonzero $\Lambda $ should reach the value
\begin{equation}
\label{eq30}
\omega = - 1
\end{equation}
in case of quintessential models it should be of the range
\begin{equation}
\label{eq31}
\omega = ( - 1; - 1 / 3)
\end{equation}
and in our ENU model \cite{sim0105090}
\begin{equation}
\label{eq32}
\omega = - 1 / 3
\end{equation}
It is worth mentioning, however, that to obtain exact value of $\omega $, 
the exact value of Hubble constant must be known.

Summarizing the conclusions offered in the present contribution it should be 
pointed out that in the majority of conventional models it is postulated 
that $E_{CBR} \approx T_{CBR} \approx a^{ - 1}$, $\varepsilon _{CBR} \approx 
a^{ - 4},S = \mbox{const}$. 

In the ENU, $E_{CBR} \approx T_{CBR} \approx a^{ - 3 / 4}$, $\varepsilon 
_{CBR} \approx a^{ - 3},S \approx a^{ - 1 / 4}$. Observations are in accord 
with the values derived by ENU model. Other models explain the above 
mentioned observed relations as a consequence of the matter emerging from 
behind event horizon due to the Universe expansion deceleration. The latest 
measurements, however, suggest that the Universe expansion might accelerate 
and exclude its deceleration. In the ENU the expansion neither decelerates 
nor accelerates, it is constant and equal to the velocity of light.

\end{document}